\documentclass[aps,prl,reprint,groupedaddress]{revtex4-1}

\usepackage{graphicx}
\usepackage{dcolumn}
\usepackage{bm}
\usepackage{hyperref}
\usepackage{amsmath}

\newcommand{\ab}{\bar a}
\newcommand{\ah}{\hat a}
\newcommand{\as}{\alpha_s}

\newcommand{\nn}{\nonumber}

\newcommand{\eqn}[1]{(\ref{#1})}

\newcommand{\tvs}{\vbox{\vskip 6mm}}
\newcommand{\MSb}{{\overline{\rm MS}}}
\newcommand{\sfrac}[2]{\mbox{$\frac{#1}{#2}$}}

\begin{document}


\title{Absence of even-integer \boldmath{$\zeta$}-function values in 
Euclidean physical quantities in QCD}


\author{Matthias Jamin}
\affiliation{IFAE, BIST, Campus UAB, 08193 Bellaterra (Barcelona) Spain}
\affiliation{ICREA, Pg.~Llu\'\i s Companys 23, 08010 Barcelona, Spain}

\author{Ramon Miravitllas}
\affiliation{IFAE, BIST, Campus UAB, 08193 Bellaterra (Barcelona) Spain}

\date{\today}

\begin{abstract}
At order $\as^4$ in perturbative quantum chromodynamics, even-integer
$\zeta$-function values are present in Euclidean physical correlation functions
like the scalar quark correlation function or the scalar gluonium correlator.
We demonstrate that these contributions cancel when the perturbative expansion
is expressed in terms of the so-called $C$-scheme coupling $\hat\alpha_s$ which
has recently been introduced in Ref.~\cite{bjm16}. It is furthermore conjectured
that a $\zeta_4$ term should arise in the Adler function at order $\as^5$
in the $\MSb$-scheme, and that this term is expected to disappear in the
$C$-scheme as well.
\end{abstract}


\maketitle

\section{Introduction}

In the past, it has been noted several times that even-integer values of the
Riemann $\zeta$-function are absent in the perturbative expansion of some
Euclidean physical quantities in quantum chromodynamics (QCD). One prominent
example is the Adler function up to order $\as^4$ \cite{bck08}, and explanations
for this behaviour were provided in the literature \cite{bck17,bc10,bro99}.
However, the regularity for example fails in the scalar quark correlation
function \cite{bck06} and the scalar gluonium correlator \cite{hruvv17}, both
also being analytically available up to order $\as^4$ in the $\MSb$-scheme
\cite{bbdm78}.

In Ref.~\cite{bjm16}, together with D.~Boito, we introduced a novel
definition of the QCD coupling, $\hat\alpha_s$, which reflects the simple
scheme-transformation properties of the $\Lambda$-parameter, such that the
scheme dependence of the coupling $\hat\alpha_s$ could be parametrised by a
single parameter $C$. Hence, $\hat\alpha_s$ was named the ``$C$-scheme''
coupling. It was furthermore found that the corresponding $\beta$-function is
then scheme independent, only depending on the first two $\beta$-coefficients
$\beta_1$ and $\beta_2$ in a way that has already been studied previously in
a different context \cite{byz92}.

In this work, we shall demonstrate that both, the Euclidean physical scalar
correlation function, as well as the scalar gluonium correlator, up to order
$\as^4$ are free of even-integer $\zeta$-function values when they are
appropriately expressed in terms of the $C$-scheme coupling. We will also give
additional arguments, why even-integer $\zeta$-function values have not yet
appeared in the Adler function, but we conjecture that a $\zeta_4\equiv\zeta(4)$
term should appear at order $\as^5$ in the $\MSb$-scheme. We furthermore
conjecture that it should again cancel when the Adler function is expressed in
$\hat\alpha_s$ and that the same might hold true for all Euclidean physical
quantities in QCD, possibly even for those in quantum field theory in general.

Our article is organised as follows: to begin we collect the required relations
for the $C$-scheme coupling. Then we briefly discuss the Adler function and
argue why a $\zeta_4$ term is only expected at order $\as^5$. The scalar quark
and gluonium correlators are presented in more detail and it is demonstrated
how even-integer $\zeta$-function values cancel after rewriting them in the
$C$-scheme coupling $\hat\alpha_s$.

\section{The \boldmath{$C$}-scheme coupling}

The $C$-scheme coupling $\ah_Q\equiv\hat\alpha_s(Q)/\pi$ has been introduced
in Ref.~\cite{bjm16}, and is defined by the relation
\begin{eqnarray}
\label{ahat}
\frac{1}{\ah_Q} + \frac{\beta_2}{\beta_1} \ln\ah_Q - \frac{\beta_1}{2}\,C
\,&\equiv&\, \beta_1 \ln\frac{Q}{\Lambda} \nn \\
&& \hspace{-16mm} \,\equiv\, \frac{1}{a_Q} + \frac{\beta_2}{\beta_1}\ln a_Q -
\beta_1 \!\int\limits_0^{a_Q}\,\frac{{\rm d}a}{\tilde\beta(a)} \,,
\end{eqnarray}
where $C$ is a free parameter,
\begin{equation}
\frac{1}{\tilde\beta(a)} \,\equiv\, \frac{1}{\beta(a)} - \frac{1}{\beta_1 a^2}
+ \frac{\beta_2}{\beta_1^2 a} \,,
\end{equation}
and both, the QCD $\Lambda$ parameter, and the coupling $a_Q$ on the right-hand
side are in a conventional renormalisation scheme, like for example the $\MSb$
scheme. The coupling $\ah_Q$ was selected such as to mimic the simple
scheme-transformation properties of the $\Lambda$ parameter.

From Eq.~\eqn{ahat} it is an easy matter to derive the corresponding
$\beta$-function, which was found to only depend on the scheme-invariant
coefficients $\beta_1=11/2-N_f/3$ and $\beta_2=51/4-19/12\,N_f$, $N_f$ the
number of quark flavours:
\begin{equation}
\label{CRGE}
-\,Q\,\frac{{\rm d}\ah_Q}{{\rm d}Q} \,\equiv\, \hat\beta(\ah_Q) \,=\,
\frac{\beta_1 \ah_Q^2}{\left(1 - \sfrac{\beta_2}{\beta_1}\, \ah_Q\right)}
\,=\, -\,2\,\frac{{\rm d}\ah_Q}{{\rm d}C} \,.
\end{equation}
Additionally, also $C$-evolution is governed by the same $\beta$-function.
This implies that in the $\ah_Q$ coupling, scale and scheme variation can
be considered on an equal footing.

Hence, transforming from a general scheme $a_Q$ to $\ah_Q$ can be performed
in two steps. From Eq.~\eqn{ahat} at $C=0$, defining $\ab_Q\equiv\ah_Q^{C=0}$,
one finds the relation
\begin{equation}
\label{afunab}
\begin{split}
a_Q \,&=\, \ab_Q + \biggl( \frac{\beta_3}{\beta_1} -
\frac{\beta_2^2}{\beta_1^2} \biggr) \ab_Q^3 + \biggl(
\frac{\beta_4}{2\beta_1} - \frac{\beta_2^3}{2\beta_1^3} \biggr) \ab_Q^4 \\
&+ \biggl( \frac{\beta_5}{3\beta_1} - \frac{\beta_2\beta_4}{6\beta_1^2} +
\frac{5\beta_3^2}{3\beta_1^2} - \frac{3\beta_2^2\beta_3}{\beta_1^3} +
\frac{7\beta_2^4}{6\beta_1^4} \biggr) \ab_Q^5 + \ldots
\end{split}
\end{equation}
Then, in a second step, the $C$ evolution can be employed to transform from
$\ab_Q$ to the general $C$-scheme coupling:
\begin{eqnarray}
\label{abfunah}
\ab_Q \,&=&\, \ah_Q + \frac{\beta_1}{2}\,C\,\ah_Q^2 + \biggl(
\frac{\beta_2}{2}\,C + \frac{\beta_1^2}{4}\,C^2 \biggr) \ah_Q^3 \nn \\
\tvs
&+& \!\biggl( \frac{\beta_2^2}{2\beta_1}\,C + \frac{5\beta_1\beta_2}{8}\,C^2
+ \frac{\beta_1^3}{8}\,C^3 \biggr) \ah_Q^4 \\
\tvs 
&+& \!\biggl( \frac{\beta_2^3}{2\beta_1^2}\,C + \frac{9\beta_2^2}{8}\,C^2 +
\frac{13\beta_1^2\beta_2}{24}\,C^3 + \frac{\beta_1^4}{16}\,C^4 \biggr) \ah_Q^5
+ \ldots \nn
\end{eqnarray}
This summarises all required relations for the $C$-scheme coupling $\ah_Q$.

\section{The Adler function}

The resummed perturbative Adler function $D(Q^2)$, $Q^2\!>\!0$, which results
from the $Q^2$-derivative of the vector correlator, and is a physical, scale-
and scheme-independent quantity, assumes the simple expression
\begin{equation}
\label{DaQ}
D(Q^2) \,=\, \frac{N_c}{12\pi^2} \sum\limits_{n=0}^\infty
c_{n,1}\,a_Q^n \,.
\end{equation}
The independent coefficients $c_{n,1}$ are known analytically up to order
$\as^4$ \cite{bck08}. Further details regarding our notation and additional
references can be found in Ref.~\cite{bj08}. For definiteness, but to keep the
expressions simple, we will only quote results for $N_f=3$. At $N_c=3$ and in
the $\MSb$-scheme, the $c_{n,1}$ were found to be:
\begin{eqnarray}
\label{cn1}
c_{0,1} &=& c_{1,1} \,=\, 1 \,, \quad
c_{2,1} \,=\, \sfrac{299}{24} - 9\zeta_3 \,, \nn \\
\tvs
c_{3,1} &=& \sfrac{58057}{288} - \sfrac{779}{4}\zeta_3 + \sfrac{75}{2} \zeta_5
\,, \\
\tvs
c_{4,1} &=& \sfrac{78631453}{20736} - \sfrac{1704247}{432}\zeta_3 +
\sfrac{4185}{8}\zeta_3^2 + \sfrac{34165}{96}\zeta_5 - \sfrac{1995}{16}\zeta_7
\,, \nn
\end{eqnarray}
where $\zeta_n\equiv\zeta(n)$. As is seen explicitly, up to order $\as^4$,
even in the $\MSb$-scheme, the $c_{n,1}$ only contain the odd $\zeta$-function
values $\zeta_3$, $\zeta_5$ and $\zeta_7$.

Next, we transform the Adler function into the $C$-scheme coupling $\ab_Q$
at $C=0$. The corresponding expansion assumes the form
\begin{equation}
\label{DabQ}
D(Q^2) \,=\, \frac{N_c}{12\pi^2} \sum\limits_{n=0}^\infty
\bar c_{n,1}\,\bar a_Q^n \,.
\end{equation}
Employing Eq.~\eqn{afunab}, only the coefficients $\bar c_{3,1}$ and
$\bar c_{4,1}$ turn out different from the $\MSb$ coefficients, and read:
\begin{eqnarray}
\label{cb3cb4}
\bar c_{3,1} &=& \sfrac{262955}{1296} - \sfrac{779}{4}\zeta_3 +
                 \sfrac{75}{2} \zeta_5 \,, \\
\tvs
\bar c_{4,1} &=& \sfrac{357259199}{93312} - \sfrac{1713103}{432}\zeta_3 +
                 \sfrac{4185}{8}\zeta_3^2 + \sfrac{34165}{96}\zeta_5 -
                 \sfrac{1995}{16}\zeta_7 \,. \nn
\end{eqnarray}
Like $c_{3,1}$ and $c_{4,1}$, the coefficients $\bar c_{3,1}$ and $\bar c_{4,1}$
still only include odd-integer $\zeta$ values up to $\zeta_7$, because the
transformation \eqn{afunab} only includes $\beta$-coefficients up to $\beta_4$,
which have $\zeta_3$ as the sole irrational component. This changes at order
$\as^5$, since $\beta_5$ also contains a $\zeta_4$ term \cite{bck16,hruvv17b}.
However, below we shall demonstrate that the $\zeta_4$ term in $\beta_5$
precisely cancels a corresponding term in the $\MSb$ coefficients of scalar
quark and gluonium correlators, such as to make the $C$-scheme coefficients
independent of $\zeta_4$.

Therefore, we conjecture that the same should also happen for the Adler
function: we suspect that the coefficient $c_{5,1}$ will contain a $\zeta_4$
term which cancels against the corresponding term in $\beta_5$ once the Adler
function is reexpressed in the $C$-scheme coupling $\ah_Q$. Under this
assumption, we can predict the component of $c_{5,1}$ which is proportional
to $\zeta_4$. At $N_c=3$, but for arbitrary number of quark flavours $N_f$,
it is found to be
\begin{equation}
\label{c5z4}
c_{5,1}^{\zeta_4} \,=\, \Big( \sfrac{2673}{512} - \sfrac{1627}{4608}\,N_f +
\sfrac{809}{6912}\,N_f^2 \Big) \zeta_4 \,.
\end{equation}

\section{The scalar quark correlator}

In the case of the scalar quark correlator, the Euclidean physical quantity
is given by the second derivative of the correlation function $\Psi(Q^2)$.
Its definition and further details can be found in Ref.~\cite{jm16}.
Furthermore, the scalar current has an anomalous dimension which is inverse
to that of the quark mass. Hence, a scale- and scheme-invariant correlator
can be obtained by multiplying with two powers of a generic quark mass
$m_Q\equiv m(Q)$. The physical scalar correlator then takes the form
\begin{equation}
\label{Psipp}
\Psi^{''}\!(Q^2) \,=\, \frac{N_c}{8\pi^2}\,\frac{m_Q^2}{Q^2}\,\biggl\{
\, 1 + \sum\limits_{n=1}^\infty \,d_{n,1}^{\,''} \,a_Q^n \,\biggr\} \,,
\end{equation}
where both, the running quark mass, as well as the running QCD coupling are to
be evaluated at the renormalisation scale $Q$. At $N_f=N_c=3$, the perturbative
coefficients $d_{n,1}^{\,''}$ take the explicit values \cite{bck06}
\begin{eqnarray}
\label{dppn}
d_{1,1}^{\,''} &=& \sfrac{11}{3} \,, \quad
d_{2,1}^{\,''} \,=\, \sfrac{5071}{144} - \sfrac{35}{2}\zeta_3 \,, \nn \\
\tvs
d_{3,1}^{\,''} &=& \sfrac{1995097}{5184} - \sfrac{65869}{216}\zeta_3 -
\sfrac{5}{2}\zeta_4 + \sfrac{715}{12} \zeta_5
\,, \\
\tvs
d_{4,1}^{\,''} &=& \sfrac{2361295759}{497664} - \sfrac{25214831}{5184}\zeta_3 +
\sfrac{192155}{216}\zeta_3^2 - \sfrac{14575}{576}\zeta_4 \nn \\
\tvs
&+& \sfrac{59875}{108}\zeta_5 - \sfrac{625}{48}\zeta_6 -
\sfrac{52255}{256}\zeta_7 \,. \nn
\end{eqnarray}
It is observed that in this case $d_{3,1}^{\,''}$ contains a $\zeta_4$ term
and $d_{4,1}^{\,''}$ both $\zeta_4$ and $\zeta_6$.

For the ensuing discussion it will be essential to remove the running effects
of the quark mass from the remaining perturbative series. This can be achieved
by rewriting the running mass $m_Q$ in terms of an invariant quark mass
$\hat m$ which is defined through the relation
\begin{equation}
\label{mhat}
m_Q \,\equiv\, \hat m \,[\alpha_s(Q)]^{\gamma_m^{(1)}/\beta_1}
\exp\Biggl\{ \int\limits_0^{a_Q} \!{\rm d}a \biggl[
\frac{\gamma_m(a)}{\beta(a)} - \frac{\gamma_m^{(1)}}{\beta_1 a} \biggr]
\Biggr\} \,,
\end{equation}
where $\gamma_m(a)$ is the quark-mass anomalous dimension and $\gamma_m^{(1)}$
its first coefficient. Accordingly, we define a modified perturbative expansion
with new coefficients $r_n$,
\begin{equation}
\label{Psippmhat}
\Psi^{''}\!(Q^2) \,=\, \frac{N_c}{8\pi^2}\,
\frac{\hat m^2}{Q^2} \,[\alpha_s(Q)]^{2\gamma_m^{(1)}/\beta_1}
\biggl\{\, 1 + \sum_{n=1}^\infty \, r_n \,a_Q^n \,\biggr\} \,,
\end{equation}
which now contain contributions from the exponential factor in eq.~\eqn{mhat}.
The order $\as^4$ coefficient $r_4$ depends on $\beta$-function coefficients
as well as quark-mass anomalous dimensions up to five-loops \cite{bck14}.
Let us remark that the $\zeta_4$ term that is present in $d_{3,1}^{\,''}$,
as well as the $\zeta_6$ term in $d_{4,1}^{\,''}$, are cancelled by the
additional contribution, while $\zeta_4$ still remains in $r_4$. The respective
cancellations have also been observed in ref.~\cite{bck17} for a related
quantity.

As the last step, similarly to the Adler function, we reexpress the QCD
coupling in terms of $\bar\alpha_s$. The perturbative expansion of $\Psi^{''}$
then assumes the form
\begin{equation}
\label{Psippabar}
\Psi^{''}\!(Q^2) \,=\, \frac{N_c}{8\pi^2}\,
\frac{\hat m^2}{Q^2} \,[\bar\alpha_s(Q)]^{2\gamma_m^{(1)}/\beta_1}
\biggl\{\, 1 + \sum_{n=1}^\infty \, \bar r_n \,\ab_Q^n \,\biggr\} \,,
\end{equation}
defining the coefficients $\bar r_n$, which take the particular values
\begin{eqnarray}
\label{rb1torb4}
\bar r_1 &=& \sfrac{442}{81} \,, \qquad
\bar r_2 \,=\, \sfrac{2510167}{52488} - \sfrac{335}{18} \zeta_3 \,, \nn \\
\tvs
\bar r_3 &=& \sfrac{12763567259}{25509168} - \sfrac{673561}{1944} \zeta_3 +
\sfrac{18305}{324} \zeta_5 \,, \\
\tvs
\bar r_4 &=& \sfrac{49275071521973}{8264970432} - \sfrac{10679302931}{1889568}
\zeta_3 + \sfrac{601705}{648} \zeta_3^2 \nn \\
\tvs
&+& \sfrac{117947335}{209952} \zeta_5 - \sfrac{3285415}{20736} \zeta_7 \,. \nn
\end{eqnarray}
As has already been pointed out above, now even the $\zeta_4$ term remaining in
$r_4$ got cancelled by a corresponding contribution in $\beta_5$, originating
from the global $\bar\alpha_s$ prefactor, such that only odd-integer
$\zeta$-function contributions persist. Even though we have just provided
results for $N_f=3$, we have convinced ourselves that the cancellation of even
$\zeta$ values does in fact take place for an arbitrary number of flavours and
a general gauge group. Furthermore, since the transformation \eqn{abfunah} only
contains the $\beta$-function coefficients $\beta_1$ and $\beta_2$ which are
rational, the absence of even $\zeta$ values also remains true for a general
$C$-scheme coupling $\hat\alpha_s$. It is hence a scheme-independent statement.

\section{The scalar gluonium correlator}

A basic two-point correlation function that is relevant for the study of
scalar gluonium can be defined as
\begin{equation}
\label{TG2G2}
\Pi_{G^2}(q^2) \,\equiv\, i\!\int\!{\rm d}x\,{\rm e}^{iqx}\,
\langle 0|T\{J_G(x) J_G(0)\}|0 \rangle \,,
\end{equation}
where the gluonic current is represented by
$J_G(x)\equiv G_{\mu\nu}^{\,a}(x)\,G^{\mu\nu\,a}(x)$ and $G_{\mu\nu}^{\,a}(x)$
is the QCD field-strength tensor.

In order to be able to define a physical quantity, one should work with a
renormalisation group invariant current. In the chiral limit, where the
operator $J_G(x)$ does not mix with $m\,\bar q(x)\,q(x)$ or $m^4$, such
a current can be chosen to be
\begin{equation}
\label{JGtilde}
\hat J_G(x) \,\equiv\, \frac{\beta(a)}{\beta_1 a}\,J_G(x) \,=\,
\frac{\beta(a)}{\beta_1 a}\,G_{\mu\nu}^{\,a}(x)\,G^{\mu\nu\,a}(x) \,.
\end{equation}
In analogy to $\Pi_{G^2}(q^2)$, we can then define the two-point correlator
for the current $\hat J_G(x)$, which expressed in terms of $\Pi_{G^2}(q^2)$
takes the form ($Q^2=-q^2$):
\begin{equation}
\label{TG2G2tilde}
\hat\Pi_{G^2}(Q^2) \,=\, \left(\frac{\beta(a_Q)}{\beta_1 a_Q}\right)^{\!2}
\Pi_{G^2}(Q^2) \,.
\end{equation}

A Euclidean physical quantity in analogy to the Adler function can be obtained
by taking three derivatives of $\hat\Pi_{G^2}(Q^2)$, leading to the definition
\begin{equation}
\label{DG2}
D_{G^2}(Q^2) \,\equiv\, -\,Q^2\,\frac{{\rm d}^3
\hat\Pi_{G^2}(Q^2)}{{\rm d}(Q^2)^3} \,.
\end{equation}
The corresponding perturbative expansion then takes the following general
form \cite{jam12}:
\begin{equation}
\label{DG2aQ}
D_{G^2}(Q^2) \,=\, \frac{(N_c^2-1)}{2\pi^2}\,a_Q^2 \sum\limits_{n=0}^\infty
g_n a_Q^n \,.
\end{equation}
Up to order $\as^4$, the coefficients $g_n$ can be extracted from the results
provided in Ref.~\cite{hruvv17}. Again at $N_f=N_c=3$, they are obtained as
follows:
\begin{eqnarray}
\label{gn}
g_0 &=& 1 \,, \quad g_1 \,=\, \sfrac{104}{9} \,, \quad
g_2 \,=\, \sfrac{87605}{648} - \sfrac{465}{8}\zeta_3 \,, \nn \\
\tvs
g_3 &=& \sfrac{52031155}{31104} - \sfrac{216701}{144}\zeta_3 +
        \sfrac{10205}{24} \zeta_5 \,, \\
\tvs
g_4 &=& \sfrac{33122537939}{1492992} - \sfrac{1833382667}{62208}\zeta_3 +
\sfrac{264275}{64}\zeta_3^2 \nn \\
\tvs
&+& \sfrac{1335}{128}\zeta_4 + \sfrac{1478075}{128}\zeta_5 -
\sfrac{2016175}{576}\zeta_7 \,. \nn
\end{eqnarray}
As anticipated above, the coefficient $g_4$ in the $\MSb$ scheme contains a
$\zeta_4$ term. Like for the scalar correlator, this is due to the anomalous
dimension of the scalar gluonium current, which leads to the global factor
$\as^2$, multiplying the correlator. As an aside, we also remark that the
leading-$N_f$ contributions are in agreement with the large-$N_f$ results
derived in Ref.~\cite{jam12}.

As in the two examples above, we conclude by rewriting the perturbative
series in terms of the $C$-scheme coupling $\ab_Q$:
\begin{equation}
\label{DG2abQ}
D_{G^2}(Q^2) \,=\, \frac{(N_c^2-1)}{2\pi^2}\,\ab_Q^2 \sum\limits_{n=0}^\infty
\bar g_n \ab_Q^n \,.
\end{equation}
For this expansion, the coefficients $\bar g_n$ are found to be:
\begin{eqnarray}
\label{gbn}
\bar g_0 &=& 1 \,, \quad \bar g_1 \,=\, \sfrac{104}{9} \,, \quad
\bar g_2 \,=\, \sfrac{178607}{1296} - \sfrac{465}{8}\zeta_3 \,, \nn \\
\tvs
\bar g_3 &=& \sfrac{20134253}{11664} - \sfrac{23979}{16}\zeta_3 +
             \sfrac{10205}{24} \zeta_5 \,, \\
\tvs
\bar g_4 &=& \sfrac{116204856235}{5038848} - \sfrac{690830641}{23328}\zeta_3 +
\sfrac{264275}{64}\zeta_3^2 \nn \\
\tvs
&+& \sfrac{59594845}{5184}\zeta_5 - \sfrac{2016175}{576}\zeta_7 \,. \nn
\end{eqnarray}
As expected, once again the $\zeta_4$ term in $g_4$ has been cancelled by the
corresponding contribution in $\beta_5$. Also in this case we have verified
that the respective cancellation is independent of the number of flavours $N_f$
and the gauge group.

\section{Conclusions}

In this work, we have demonstrated that the Euclidean physical correlation
functions corresponding to the scalar quark and scalar gluonium correlator
do not contain even-integer $\zeta$-function values in their perturbative
coefficients up to the presently analytically available order $\alpha_s^4$
when the perturbative expansion is performed in terms of the $C$-scheme
coupling $\hat\alpha_s$ \cite{bjm16}. We have shown this explicitly for the
coupling $\bar\alpha_s$ at $C=0$, but the statement remains true for an
arbitrary $C$ since the relation \eqn{abfunah} only contains $\beta_1$ and
$\beta_2$ which are rational numbers.

In the case of the Adler function, even the perturbative coefficients
in the $\MSb$ scheme up to order $\alpha_s^4$ do not contain even-integer
$\zeta$-function values. This is related to the fact that the vector current
has no anomalous dimension, and hence no prefactor depending on $\alpha_s$
arises. It is conjectured, that a $\zeta_4$ term will appear in the order
$\alpha_s^5$ coefficient $c_{5,1}$. Assuming that this term is again cancelled
in the $C$-scheme by a corresponding term in the $\beta$-function coefficient
$\beta_5$, we predict the respective component $c_{5,1}^{\zeta_4}$ proportional
to $\zeta_4$ for $\MSb$ in eq.~\eqn{c5z4}.

To our knowledge, at this moment, the cancellation of even-integer
$\zeta$-function values for perturbative expansions of Euclidean physical
correlators in the $C$-scheme coupling $\hat\alpha_s$ can only be checked
for the scalar quark and scalar gluonium correlation functions, as only these
functions are available up to the required order $\alpha_s^4$. Nonetheless,
we conjecture that the same cancellation should also take place for other
quantities. It will be exciting to see if this claim is indeed confirmed in
the future.

As a last remark, we note that compared to the Adler function coefficients,
the ones for scalar quark and gluonium correlators are substantially larger.
Already in Ref.~\cite{jm16}, we showed how the $C$-scheme coupling can be
employed in order to improve the expansion for the scalar quark correlator.
In future work, we plan to also return to this question for the scalar gluonium
correlator and furthermore intend to demonstrate how the $C$-scheme coupling
$\hat\alpha_s$ can be utilised for constructing models for the Borel transforms
of the investigated correlators, along the lines of Ref.~\cite{bj08}.

\vspace{2mm}
\begin{acknowledgments}
MJ gratefully acknowledges helpful discussions with Kostja~Chetyrkin.
The work of MJ and RM has been supported in part by MINECO Grant number
CICYT-FEDER-FPA2014-55613-P, by the Severo Ochoa excellence program of MINECO,
Grant SO-2012-0234, and Secretaria d'Universitats i Recerca del Departament
d'Economia i Coneixement de la Generalitat de Catalunya under Grant 2014 SGR
1450.
\end{acknowledgments}

\section{Note added in proof}

Meanwhile, the cancellation of even-integer $\zeta$-function values has been
demonstrated for a substantial number of additional
quantities by Davies and Vogt in Ref.~\cite{dv17}, as well as by Chetyrkin
in unpublished work (see footnote 2 in Ref.~\cite{rhuvv18}).


%

\end{document}